# Phonons, Nature of Bonding and Their Relation to Anomalous Thermal Expansion Behavior of $M_2O$ (M=Au, Ag, Cu)

M. K. Gupta, R. Mittal and S. L. Chaplot

*Solid State Physics Division, Bhabha Atomic Research Centre, Mumbai 400085, India*

S. Rols

*Institut Laue-Langevin, BP 156, 38042 Grenoble Cedex 9, France*

We report a comparative study of the dynamics of $Cu_2O$, $Ag_2O$ and $Au_2O$ (i.e. $M_2O$ with M = Au, Ag and Cu) using first principle calculations based on the density functional theory. Here for the first time we show that the nature of chemical bonding and open space in the unit cell are directly related to the magnitude of thermal expansion coefficient. A good match between the calculated phonon density of states and that derived from inelastic neutron scattering measurements is obtained for $Cu_2O$ and $Ag_2O$. The calculated thermal expansions of $Ag_2O$ and $Cu_2O$ are negative, in agreement with available experimental data, while it is found to be positive for $Au_2O$. We identify the low energy phonon modes responsible for this anomalous thermal expansion. We further calculate the charge density in the three compounds and find that the magnitude of the ionic character of the $Ag_2O$, $Cu_2O$, and $Au_2O$ crystals is in decreasing order, with an Au-O bond of covalent nature strongly rigidifying the $Au_4O$ tetrahedral units. The nature of the chemical bonding is also found to be an important ingredient to understand the large shift of the phonon frequencies of these solids with pressure and temperature. In particular, the quartic component of the anharmonic term in the crystal potential is able to account for the temperature dependence of the phonon modes.

# I. INTRODUCTION

Thermal expansion is an essential consideration for the possible applications of a compound. During the last two decades, so called "anomalous thermal expansions" have been reported in many frame-work solids. An "anomalous thermal expansion" characterizes the capability of a solid to contract upon heating i.e having a negative thermal expansion (NTE) coefficient. The microscopic understanding of the mechanism behind NTE is an important task in the quest of engineering composite materials with a controled thermal expansion. In the last years, many experimental and theoretical studies have been reported on NTE materials [1-16] e.g. on $ZrW_2O_8$, $ZrV_2O7$, $HfV_2O_7$, $ScF_3$. In particular, we have identified [12-16] the energy regime of the phonons responsible for the NTE behaviour of these compounds and gave insight into the possible microscopic mechanism.

The compounds $M_2O$ (M=Ag, Cu and Au) [17,18] crystallize in a simple cubic lattice (space group Pn-3m). The M atoms are linearly coordinated by two oxygen atoms, while oxygen is tetrahedrally coordinated by M atoms. $Ag_2O$ shows a large isotropic negative thermal expansion (NTE) over its entire temperature range of stability, i.e. up to ~ 500 K, while $Cu_2O$ only shows a small NTE below room temperature. At the moment, no experimental data are found in the literature for $Au_2O$, but we anticipate that our simulations on this isostructural system will generate effort in this direction. Extensive experimental data including specific heat measurements [23, 24], Raman and IR detection of long-wavelength optically active phonons [25, 26] together with neutron derived phonon dispersion relations and phonon density of states [27, 28] have been reported for $Cu_2O$ while the calculations of the volume thermal expansion in $Ag_2O$ and $Cu_2O$ have also been published [15, 16, 28]. EXAFS measurements on $Ag_2O$ and $Cu_2O$ [19–22] indicate that the mechanism at the origin of their NTE involves deformations of the $M_4O$ tetrahedral units (M = Ag, Cu), rather than simple rigid units vibrations. Also it has been suggested [22] that the large difference between the NTE coefficient of $Ag_2O$ and $Cu_2O$ not only originates from a mass effect but also from differences from the chemical interaction between the atoms

Measurements of temperature and pressure dependence of the phonon modes provide information about their anharmonic nature. Such investigations showed that for $Zn(CN)_2$ [16, 17] and cubic $ReO_3$ [29], the modes which reacted strongly with temperature contributed largely to the NTE of the compounds. However phonons showing anharmonicty with temperature in $ZrW_2O_8$ are identified [14] to be not necessarily those showing anharmonicity with pressure.

In our previous studies, we have reported [16] ab-initio calculations and neutron measurements of the phonon density of states of $Ag_2O$. The temperature dependence of the spectrum indicates [16] a strong anharmonic nature of phonon modes with frequencies around 2.4 meV. In the present paper, we extend this study to the case of $Au_2O$ and $Cu_2O$ and further calculate the temperature and pressure dependence of the phonon modes for all three compounds. Much insight into the physics at play in the NTE properties of these systems can be derived from a systematic study of the series Cu, Ag, Au. This implies using the same simulation technique and the same experimental apparatus. We find that $Ag_2O$ and $Cu_2O$ have a NTE while we find that $Au_2O$ has a large positive expansion with increasing temperature. In the next sections, we discuss the origins for these large differences in the amplitude and sign of the thermal expansion coefficients between these isostructural compounds. Using the density functional perturbation approach we calculate the Born effective charge dielectric constant, elastic constants and other properties. We find that $Au_2O$ has a less ionic character than $Ag_2O$ and $Cu_2O$ and discuss the consequences of tighter Au-O bonds on the lattice expansion and anharmonicity of the crystal. We also present the calculated temperature dependence of low energy phonons modes at the $\Gamma$, X, M and R-points in all three $M_2O$ compounds which allow us to evaluate the relative weight between implicit and explicit anharmonicity.

## II. EXPERIMENTAL

Polycrystalline samples of $Cu_2O$ and $Ag_2O$ (99.9% purity purchased from Sigma Aldrich) were wrapped inside a thin Aluminum foil and fixed at the end of an orange cryostat stick for the measurements. The inelastic neutron scattering experiments were performed using the IN4C time of flight spectrometer at the Institut Laue-Langevin (ILL), in Grenoble, France. The measurements were done in the neutron-energy-gain mode using the incident neutron energy of 14.2 meV (2.4 Å) at 320 K. The incoherent approximation [30] was used for extracting the neutron weighted phonon density of states from the measured scattering function $S(Q,E)$.

## III. COMPUTATIONAL DETAILS

The Vienna based ab-initio simulation package (VASP) [31,22] was used for the calculations. The generalized gradient approximation (GGA) exchange correlation given by Perdew, Becke and Ernzerhof [33,34] with projected–augmented wave was chosen. The plane wave pseudo-potential with plane wave kinetic energy cutoff of 880 eV, 1000 eV and 1200 eV for $Ag_2O$, $Cu_2O$ and $Au_2O$ respectively was used. The integrations over the Brillouin zone were sampled on a 8×8×8 grid of k-points generated by the Monkhorst-pack method [35] for all three compounds. The above parameters are found to be sufficient to obtain a total energy convergence of less than 0.1 meV. The Hellman-Feynman forces were calculated by the finite displacement method (displacement 0.04 Å) using a 2×2×2 super cell. Total energies calculations were performed for the six distinct atomic configurations resulting from individual displacements of the two symmetry-inequivalent atoms (space group no. *224 Pn-3m*) along the three Cartesian directions ($\pm x$, $\pm y$ and $\pm z$). The latter calculations were performed using the fully relaxed structures as initial configurations. The convergence criteria for the total energy and ionic forces calculations were set to $10^{-8}$ eV and $10^{-5}$ eV Å$^{-1}$, respectively. The phonon properties (e.g. dispersion curves, density of states) for the three compounds were calculated using the PHONON software [36]. The calculated unit cell parameters for $Ag_2O$ and $Cu_2O$ are in agreement with the experimental data [17] (see TABLE I), while the calculated lattice parameter for $Au_2O$ is in agreement with previous calculations [37].

## IV. RESULTS AND DISCUSSION

### A. Phonon spectra

Low energy phonons are known [10-16] to be highly anharmonic for compounds exhibiting anomalous thermal expansion. Fig. 1 shows the low energy part of the calculated dispersion relation and phonon density of states in the $M_2O$ compounds (the full energy range is shown in Fig. S1 [38]). As expected for isostructural compounds, the dispersion curves below 10 meV in $Ag_2O$ and $Au_2O$ are found to be similar, while in $Cu_2O$ these modes are shifted to higher energies, which can be understood considering the smaller mass of Cu. In general, we find a very good agreement between both our calculations for $Cu_2O$ and those of Bohnen et al. [28], together with their experimental dispersion curves reported in their paper, and with the measured phonon spectra measured (Fig. 2) in this study for $Cu_2O$ and $Ag_2O$ [16]. The latter observations validate our approach using ab-initio calculations, and give us confidence into the properties we derive from them, especially those extended to $Au_2O$.

The computed partial densities of states are shown in Fig. 3. These are obtained by atomic projections of the one-phonon eigenvectors and reflect the contribution of the different atoms to the spectrum. The contributions due to Ag, Cu or Au are located below 20 meV, the lightest atom Cu (63.54 amu) having its contributions extending up to the larger frequency. The masses of Ag (107.87 amu) and Au (197.97 amu) being different, one would also expect a renormalization of the phonon frequencies. However the first two peaks in the density of states are at nearly the same energies i.e. 3 meV and 6 meV. This observation suggests that the chemical bonds in $Ag_2O$ and $Au_2O$ are of different strength. The oxygen vibrations in all the three compounds extend over the entire phonon spectral range with maximum weight for frequencies above 50 meV. The M-O stretching modes in $Cu_2O$, $Au_2O$ and $Ag_2O$ are up to 75 meV, 70 meV and 65 meV respectively. The different spectral range for these modes may reflect the different M-O bond lengths and difference in nature of M-O bonding. The smallest Cu-O bond (1.866 Å) results in shifting of energies up to the highest spectral range of 75 meV. However Ag-O and Au-O bond lengths 2.082 Å and 2.078 Å respectively in $Ag_2O$ and $Au_2O$ are similar but the stretch mode of Au is at larger frequencies. This suggests that the Au-O bond may have a more covalent nature as compared to the Ag-O bond for which an ionic nature dominates.

Finally, the lowest transverse acoustic modes in $Ag_2O$ and $Au_2O$ give rise to the first peak in the density of states at about 3 meV, clearly observed in the phonon spectra (see Fig. 2 and 3). The equivalent peak in the $Cu_2O$ spectrum is observed at 6 meV.

**B. Pressure dependence of phonon modes and thermal expansion behavior**

In the quasi-harmonic approximation the volume thermal expansion coefficient [39] of a crystalline material, is given by the following relation: $\alpha_V = \frac{1}{BV}\sum_i \Gamma_i C_{Vi}(T)$. Here $\Gamma_i = -\frac{V}{E_i}\frac{dE_i}{dV}$ is the mode Grüneisen parameter, which is a measure of the volume/pressure dependence of the phonon frequency. $C_{Vi}(T)$ is the specific heat contribution of the $i^{th}$ phonon mode (of energy $E_i$) at temperature T, while B and V stand for the volume and the bulk modulus of the compound, respectively. In the above relation, all the quantities but $\Gamma_i$ are positive at all temperatures. Therefore the sign and magnitude of $\Gamma_i$ govern the thermal expansion of the lattice, while the phonon energy range over which $\Gamma_i$ is negative determines the temperature range over which the material will show NTE.

The calculated elastic constants and bulk modulus are given in TABLE I. The experimental data [40] are only available for $Cu_2O$ which agree very well with our calculated values. The calculated pressure dependence of the phonon dispersions (see Fig. 1) shows that in case of the $Ag_2O$ and $Cu_2O$ crystals, the lowest energy modes along Γ-X-M and Γ-M line soften with pressure in contrast to the modes along the M-R and Γ-R line. The softening is found to be negligible in $Au_2O$. The low energy optic modes in $Ag_2O$ also soften in contrast to the case of $Au_2O$, where these modes harden on increasing the pressure.

The pressure dependence of the phonon spectra have been calculated in the entire Brillouin zone to allow for the calculation of the energy dependence of the Grüneisen parameter $\Gamma(E)$ (Fig. 4 and Fig. S2 [38]) and further processed to obtain the thermal expansion coefficient $\alpha_V(T)$ (Fig. 5) as discussed below. For $Ag_2O$ and $Au_2O$, the energy range for negative $\Gamma(E)$ extends up to ~3.5 meV. However the magnitude is much larger for the former compound, reaching a value of -40 for the lowest modes, while for $Au_2O$ the maximum negative $\Gamma(E)$ reaches -10. For $Cu_2O$ the phonons below 6.5 meV have negative $\Gamma(E)$ with a maximum negative value of -4.5. The latter results are completely coherent to those obtained by Bohnen et al. [28] using the same approach.

The calculated volume thermal expansion coefficient $\alpha_V(T)$ is plotted on Fig. 5 as a function of temperature. Negative thermal expansion is calculated in $Ag_2O$ over its temperature range of stability of about 500 K, while $Cu_2O$ and $Au_2O$ have negative $\alpha_V(T)$ below room temperature and below 16 K, respectively. The most negative $\alpha_V(T)$ values for $Ag_2O$ (-44 $\times 10^{-6}$ $K^{-1}$) and $Cu_2O$ (-8 $\times 10^{-6}$ $K^{-1}$) are respectively obtained at 40 K and 75K. The maximum negative value of $\alpha_V(T)$ for $Au_2O$ is much reduced compared to the other two compounds and reaches ~ -2 $\times 10^{-6}$ $K^{-1}$ at T ~ 8 K. As mentioned in the previous section, one understand the absence of NTE in the $Au_2O$ lattice as resulting from the combination of two effects: 1) reduced absolute values of negative $\Gamma(E)$ (compared to $Ag_2O$) and 2) reduced energy range for the phonon modes with negative $\Gamma(E)$ (compared to $Cu_2O$). The comparison between the available experimental data of volume thermal expansion along with our calculations is shown in Fig 6. We have also calculated (Fig. 7) contributions of phonons as a function of energy E to the volume thermal expansion at 300 K. As shown in Fig. 7 the maximum negative contribution to volume thermal expansion coefficient is from the modes of energy around 4 to 5 meV.

The nature of the low energy phonon modes contributing to the NTE can be visualized through animations [38]. The eigenvectors of a selection of them have also been plotted on Fig. 8. The lowest $\Gamma$-point optical mode corresponds to the rotation of $M_4O$ tetrahedral and the lowest X and M point modes have negative Grüneisen parameter in all the three compounds. X-point mode involves bending of M-O-M chains. The M atoms connected to various $M_4O$ have different displacements indicating significant distortion of $M_4O$ tetrahedra. This mode seems to contribute maximum to NTE in $Ag_2O$. The M-point mode involves rotation, translation as well as distortion of the $M_4O$ tetrahedra, while for R-point the amplitude of all the atoms is similar and it indicates translational motion of $M_4O$ as a rigid unit.

## C. Specific heat and mean squared thermal amplitudes

We have used the calculated total and partial phonon density of states to calculate the temperature dependence of the specific heat $C_p$ (Fig. 9) and the mean squared displacement $<u^2>$ of the atoms (Fig 10) of the three compounds. The calculated $C_p$ agrees very well with the experimental data. In particular, the sharp rise at low temperatures is correctly reproduced, which proves again that the low-energy part of the calculated phonon spectra is reliable, at least for the $Cu_2O$ and $Ag_2O$ lattices. Also, the calculated specific heats of $Ag_2O$ and $Au_2O$ are nearly same which is consistent with the similarity of the low energy part of the phonon spectra of both the compounds (Fig. 3). Also, our calculations reproduce very well the lower specific heat at low temperatures of the Cu compound, which is a consequence of the general energy up shift of the singularities in the phonon spectra of $Cu_2O$ compared to $Ag_2O$ (and $Au_2O$).

For $Ag_2O$, we calculate that both the Ag and O atoms have similar $<u^2>$ values at all T (see Fig. 10), and that these values are much larger than those calculated for the $Au_2O$ and $Cu_2O$ compounds for the same T. In particular, they are found to be twice those calculated for $Au_2O$, an effect that can easily be understood considering the mass ratio between Ag and Au for a similar density of states. We calculate that the $<u^2>$ values are the smallest for the $Cu_2O$ compound, as resulting from its phonon spectrum renormalized to higher frequencies. For comparison the experimental data [41] of $<u^2>$ of atoms in $Cu_2O$ are also shown, which are in qualitative agreement with our calculations, although with even smaller $<u^2>$.

## D. Bonding in $M_2O$ (M=Ag, Au, Cu)

It is clear that the large difference in the thermal expansion that we calculated for the three $M_2O$ compounds reflects a difference in the bonding from one compound to the other. In addition, the presence of large voids in the unit cell renders the structure even more sensitive to subtle differences in bond strength. In order to understand the nature of the M—O bonding we have calculated the charge density for the three compounds (Fig. 11). We find that the bonding character of the Ag-O bond is more ionic than that of the Cu-O bond. We find that the Au-O bond is highly directional with the charge density elongated towards the O atom *i.e.* indicating a covalent nature, as suggested by previous studies [42]. The change of bonding from an ionic to a covalent character is due to different intra-atomic hybridization between the d, s and p states [37]. The computed Born effective charges (see TABLE II) for oxygen atoms in $Ag_2O$, $Cu_2O$ and $Au_2O$ are -1.28, -1.18 and -0.54 respectively. The latter values also reflect the larger ionic character for the Ag-O bond compared to Au-O.

The compounds $Ag_2O$ and $Au_2O$ have an almost identical lattice parameter ($Ag_2O$ =4.81 Å and $Au_2O$ =4.80 Å) and similar Ag/Au-O bond lengths. However, the covalent and directional Au-O bond rigidifies the $Au_4O$ tetrahedra, making them less susceptible to distortion, bending or rotation than their $Ag_4O$ counterpart. This is revealing the microscopic origin of the large NTE in $Ag_2O$. As discussed in the previous sections, we found that $Ag_2O$ shows a large softening of its transverse acoustic modes along the Γ-X-M line with increasing pressure while in $Au_2O$ this softening is not observed. Also in case of $Ag_2O$ high energy optical modes also show softening in contrast to $Au_2O$, where these modes become hard with pressure.

Now if we compare the $Cu_2O$ and $Ag_2O$ cases, for which the nature of bonding is almost similar, we find that both compounds exhibit negative thermal expansion. However there is a large difference in the magnitude of the thermal expansion coefficient. The Cu-O (1.87 Å) bond length is much smaller than the Ag-O (2.08 Å) bond. The $Cu_4O$ tetrahedral units are therefore much more compact than $Ag_4O$, rendering distortion less favorable in $Cu_4O$ as compared to $Ag_4O$. In addition, the difference in the open space in the unit cell between the two compounds leads to differences in the magnitude of the distortions and hence difference in the NTE coefficient. Here the open structure nature of the lattice regulates the extent of the NTE.

**E. Temperature dependence of the phonon modes**

The measurement of the temperature dependence of the $Ag_2O$ phonon density of states shows [4] that the singularity at 2.4 meV in the phonon spectra at 150 K shifts to 2.8 meV on heating the sample to 310 K (such measurements are not available for $Au_2O$ and $Cu_2O$). The change in phonon energies originates from "implicit" and "explicit" anharmonicities [43]. The implicit anharmonicity reflects the unit cell volume dependence of the phonon frequency. By contrast, the explicit anharmonicity links the phonon frequencies to the amplitude of the atomic vibrations, i.e. exploring the non-harmonic part of the potential. The temperature dependent measurements include both effects. From the value of the Grüneisen parameter we could estimate that the volume compression from 150 K to 310 K would soften the 2.4 meV phonons by 0.04 meV only. Therefore, the observed hardening of the phonons with increasing temperature is an evidence for a large explicit type anharmonicity in this system.

The 2.4 meV peak in the density of states of $Ag_2O$ [4] results from the lowest energy phonon modes at the X-point and M-point in the Brillouin zone. Here we present the temperature dependence of the lowest energy phonon modes at the Γ, X, M and R-points for the three $M_2O$ compounds. The details of the perturbation method as used for our calculations are given in Ref. [44-46]. First we have computed potential energy profile (Fig. 12) for the Γ, X, M and R-point phonon modes as a function of the mode amplitude. The crystal potential energy is then fitted to the expression $V(\theta_j) = a_{0,j} + a_{2,j}\xi_j^2 + a_{3,j}\xi_j^3 + a_{4,j}\xi_j^4$ (where $\xi_j$ is the normal coordinate of the j$^{th}$ phonon mode and $a_{2,j}$, $a_{3,j}$, and $a_{4,j}$ are the coefficients of the harmonic and third and fouth order anharmonic terms, respectively). The energy of the modes may increase or decrease with increase of temperature, depending on the sign (negative or positive) of the coefficients of the anharmonic terms. The anhamonicity parameters (Table III) are obtained by fitting the above equation to the calculated energy profile and used for calculating the temperature dependence of the phonon mode frequencies.

We find that, for all compounds, the cubic component $a_{3,j}$ is zero due to the symmetry of these modes. As shown in Fig. 13 there is a significant quartic contribution in $Ag_2O$ and $Cu_2O$, while in $Au_2O$ where M-O-M bonding is covalent the quartic contribution is found to be small. The temperature dependence of a few phonon modes in $M_2O$ is shown in Fig. 13. The calculated temperature dependence of low energy X-point phonon mode in $Ag_2O$ is found to be in qualitative agreement with the temperature dependence measurements of phonon peak at 2.4 meV in the density of states [16], which involves an average over the entire Brillouin zone.

The calculated change in energy of the low energy phonon modes on increasing the temperature from 0 to 300 K together with their respective Grüneisen parameter values are given in Table IV. The Γ-point mode has a very low Grüneisen parameter for all compounds. This indicates that Γ-point modes weakly react to lattice compression or expansion. Its calculated temperature dependence is also found to be very weak in the three $M_2O$ compounds. It can be seen that the X-point mode has the maximum –negative- Grüneisen parameter among all the compounds. This mode also shows strong explicit temperature dependence for the $Ag_2O$ lattice, while the overall temperature dependence (implicit+explicit) is moderate. The M-point mode in $Ag_2O$ reacts only weakly to temperature variations, an observation also valid for $Au_2O$ and $Cu_2O$. The R-point mode have large positive Γ(E) values for all three compounds. We find that for the $Ag_2O$ lattice, the calculated temperature dependence of the mode shows a significant positive shift with temperature while it remains almost invariant with temperature in the case of $Au_2O$ and shows only moderate temperature dependence in the case of $Cu_2O$. TABLE IV shows that $Ag_2O$, with ionic bonding, reacts equally to temperature and volume changes, while $Au_2O$ with covalent bonding reacts the least to these changes.

## V. CONCLUSIONS

We have reported comparative ab-initio calculations of phonon spectra as well as thermal expansion behavior in $Au_2O$, $Ag_2O$ and $Cu_2O$. The calculations are in good agreement with the experimental inelastic neutron scattering phonon spectra. The ab-initio calculations are used for understanding the temperature and pressure dependence of the phonon modes in $M_2O$. The calculated thermal expansion behavior of these compounds is in agreement with the available experimental data. We find that although low energy phonon modes of similar energies are present in all the $M_2O$ compounds, the nature of bonding as well as open space in the unit cell are important in governing the thermal expansion behavior. The calculated quartic anharmonicity of phonons is able to account for temperature dependence of phonon modes in $M_2O$.

TABLE I. Comparison of the calculated structural parameters and elastic constants of M$_2$O (M=Ag, Au, Cu) with the experimental data. The experimental data of lattice parameters for Ag$_2$O and Cu$_2$O is at 15 K and 293 K respectively, while the calculations are performed at 0 K. The values in the brackets give the experimental data of the lattice constants [17] and elastic constants and bulk modulus[40].

|  | Calc. a (Å) | $C_{11}$(GPa) | $C_{44}$(GPa) | $C_{12}$(GPa) | B(GPa) |
|---|---|---|---|---|---|
| Ag$_2$O | 4.81 (4.745) | 80.0 | 0.6 | 70.8 | 73.8 |
| Au$_2$O | 4.80 | 101.1 | 2.4 | 94.9 | 97.1 |
| Cu$_2$O | 4.31 (4.268) | 127.2 (121) | 6.3 (10.9) | 105.4 (105) | 112.7 (112) |

TABLE. II. Calculated Born effective charges (Z) in M$_2$O (M=Ag, Au, Cu).

| Atom | $Z_{xx}$ | $Z_{xy}$ | $Z_{xz}$ | $Z_{yx}$ | $Z_{yy}$ | $Z_{yz}$ | $Z_{zx}$ | $Z_{zy}$ | $Z_{zz}$ |
|---|---|---|---|---|---|---|---|---|---|
| Ag$_2$O | | | | | | | | | |
| O | -1.28 | 0 | 0 | 0 | -1.28 | 0 | 0 | 0 | -1.28 |
| Ag | 0.64 | 0.63 | 0.63 | 0.63 | 0.64 | 0.63 | 0.63 | 0.63 | 0.64 |
| Au$_2$O | | | | | | | | | |
| O | -0.54 | 0 | 0 | 0 | -0.54 | 0 | 0 | 0 | -0.54 |
| Au | 0.27 | 0.92 | 0.92 | 0.92 | 0.27 | 0.92 | 0.92 | 0.92 | 0.27 |
| Cu$_2$O | | | | | | | | | |
| O | -1.18 | 0 | 0 | 0 | -1.18 | 0 | 0 | 0 | -1.18 |
| Cu | 0.59 | 0.44 | 0.44 | 0.44 | 0.59 | 0.44 | 0.44 | 0.44 | 0.59 |

TABLE. III. Results of anharmonic calculations. The value of the parameters '$a_{3,j}$' and '$a_{4,j}$' are extracted from fitting of equation $V(\theta_j) = a_{0,j} + a_{2,j}\xi_j^2 + a_{3,j}\xi_j^3 + a_{4,j}\xi_j^4$ to the potential of the mode with fixed value of '$a_{2,j}$' as determined from the phonon energies. The value of '$a_{3,j}$' is found to be zero.

| Wave vector | E (meV) | $a_{2,j}(\times 10^{-3})$ | $a_{4,j}(\times 10^{-6})$ | $\Delta_j^{(4)}(\times 10^{-3})$ (meV) |
|---|---|---|---|---|
| | | Ag$_2$O | | |
| Γ | 5.50 | 3.68 | 3.51 | 5.70 |
| X | 1.85 | 0.43 | 1.6 | 22.14 |
| M | 2.54 | 0.77 | 0.98 | 7.70 |
| R | 3.76 | 1.62 | 4.98 | 19.76 |
| | | Au$_2$O | | |
| Γ | 5.60 | 3.76 | -0.27 | -0.44 |
| X | 2.25 | 0.61 | 0.54 | 5.35 |
| M | 2.43 | 0.71 | 0.61 | 5.11 |
| R | 3.58 | 1.54 | 1.27 | 4.96 |
| | | Cu$_2$O | | |
| Γ | 8.46 | 8.61 | 12.13 | 8.50 |
| X | 4.48 | 2.41 | 6.97 | 17.44 |
| M | 6.30 | 4.77 | 1.90 | 2.40 |
| R | 7.65 | 7.05 | 16.90 | 14.26 |

TABLE. IV. The calculated change in energy of low-energy phonon modes on increase of temperature from 0 to 300 K. $E_i$ and $\Gamma_i$ are the energy and Grüneisen parameter of $i^{th}$ mode, respectively at 0 K. While $\Delta E_V$, $\Delta E_A$ and $\Delta E_T$ are the change in energy of mode due to change in volume (implicit anharmonicity), increase in thermal amplitudes of atoms (explicit anharmonicity) and total change in energy of mode on increase of temperature from 0 to 300 K. The parameters E, $\Delta E_V$, $\Delta E_A$ and $\Delta E_T$ are in "meV" units.

| Wave vector | $E_i$ | $\Gamma_i$ | $\Delta E_V$ | $\Delta E_A$ | $\Delta E_T$ |
|---|---|---|---|---|---|
| $Ag_2O$ | | | | | |
| Γ | 5.50 | -0.6 | -0.01 | 0.05 | 0.04 |
| X | 1.88 | -18.2 | -0.12 | 0.60 | 0.48 |
| M | 2.54 | -10.9 | -0.10 | 0.15 | 0.05 |
| R | 3.90 | 7.3 | 0.10 | 0.25 | 0.35 |
| $Au_2O$ | | | | | |
| Γ | 5.60 | 0.7 | -0.02 | 0.00 | -0.02 |
| X | 2.25 | -4.4 | 0.04 | 0.12 | 0.16 |
| M | 2.43 | -4.1 | 0.04 | 0.10 | 0.14 |
| R | 3.58 | 6.5 | -0.10 | 0.07 | -0.03 |
| $Cu_2O$ | | | | | |
| Γ | 8.46 | -1.1 | -0.01 | 0.04 | 0.03 |
| X | 4.48 | -4.4 | -0.02 | 0.18 | 0.16 |
| M | 6.30 | -4.4 | -0.03 | 0.17 | 0.14 |
| R | 7.65 | 8.6 | 0.07 | 0.08 | 0.15 |

FIG. 1. (Color online) The calculated low energy part of the phonon dispersion relation of $M_2O$ (M=Ag, Au and Cu). The Bradley-Cracknell notation is used for the high-symmetry points along which the dispersion relations are obtained. $\Gamma=(0,0,0)$; $X=(1/2,0,0)$; $M=(1/2,1/2,0)$ and $R=(1/2,1/2,1/2)$.

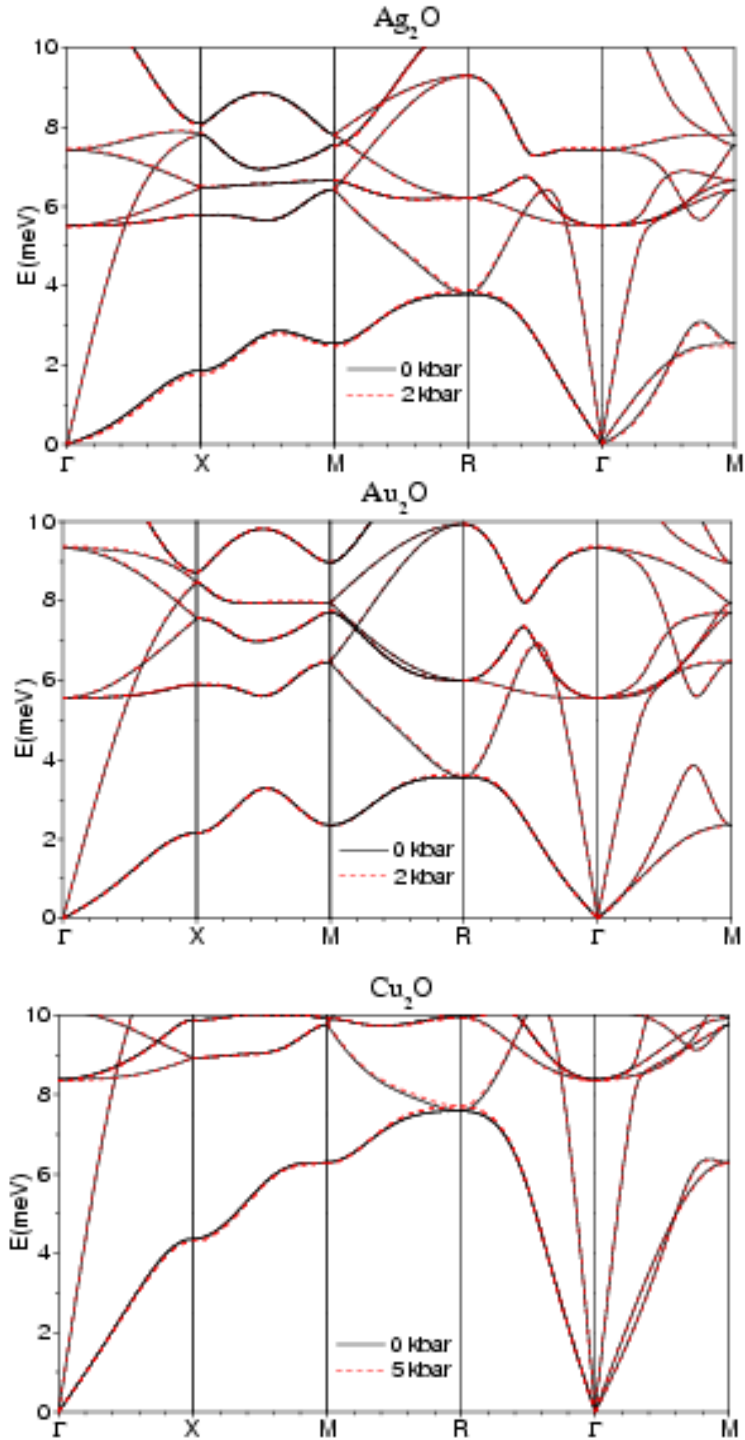

FIG. 2 Experimental (symbols plus line) and calculated (solid line) neutron-weighted phonon density of state of $M_2O$ (M=Ag, Au and Cu) compounds. The experimental phonon spectra as well as ab-initio calculation of $Ag_2O$ are already published [16] and shown here for comparison with $Au_2O$ and $Cu_2O$. The calculated spectra have been convoluted with a Gaussian of FWHM of 15% of the energy transfer in order to describe the effect of energy resolution in the experiment.

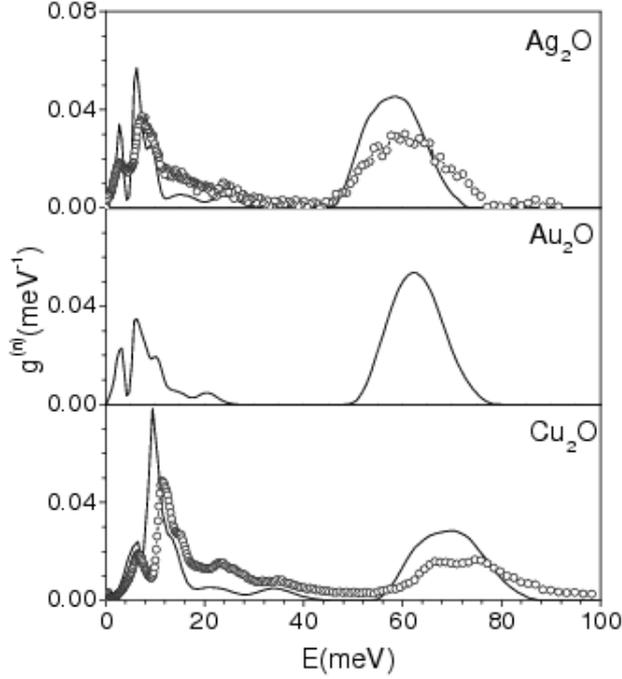

FIG. 3. (Color online) Normalized partial density of states of various atoms and total one-phonon density of states in $M_2O$ (M=Ag, Au and Cu) compounds. The calculations for $Ag_2O$ [16] are shown here for comparison with $Au_2O$ and $Cu_2O$.

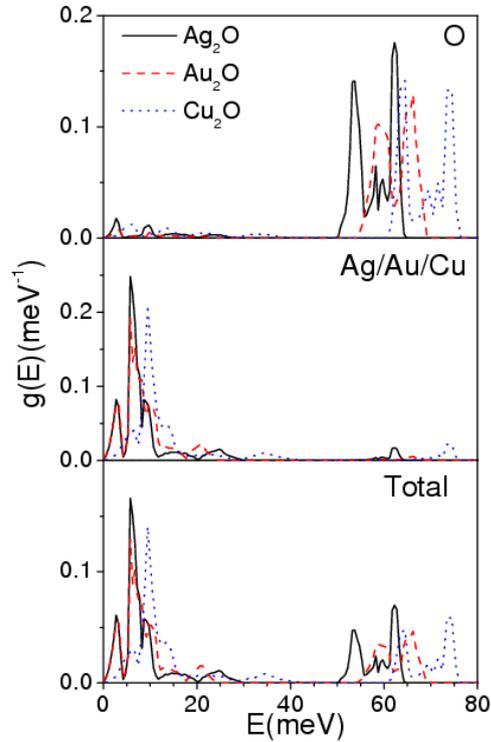

FIG. 4. (Color online) The calculated Grüneisen Parameter of $M_2O$ (M=Ag, Au and Cu). The calculations for $Ag_2O$ [16] are shown here for comparison with $Au_2O$ and $Cu_2O$.

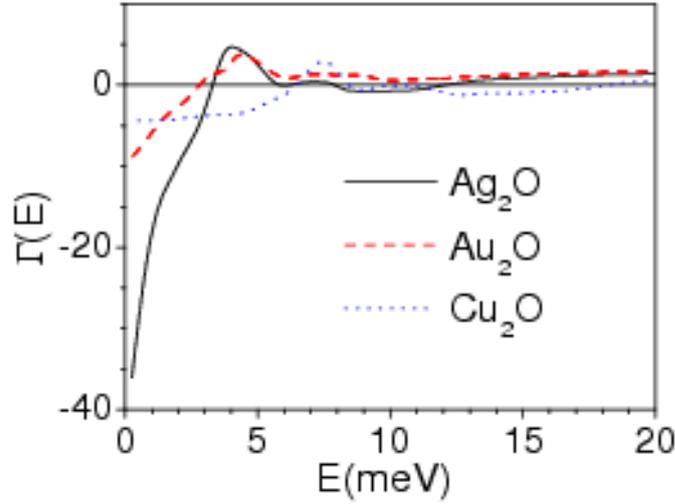

FIG. 5. (Color online) Volume thermal expansion ($\alpha_V$) coefficient as a function of temperature in $M_2O$ (M=Ag, Au and Cu). The calculations for $Ag_2O$ [16] are shown here for comparison with $Au_2O$ and $Cu_2O$.

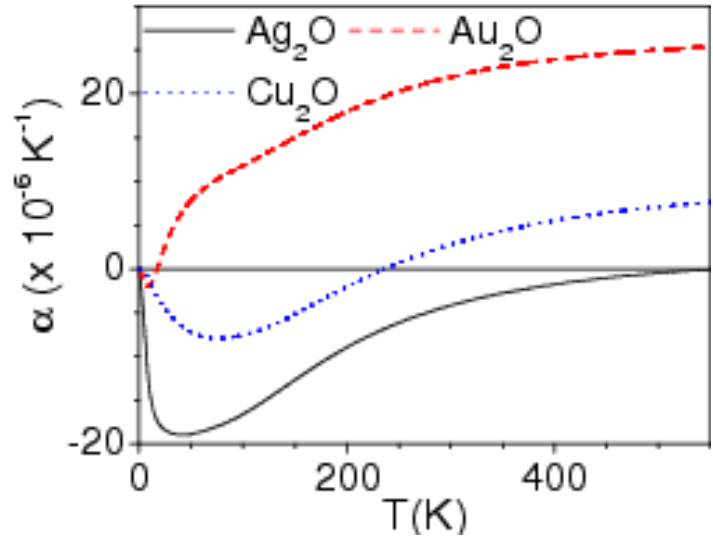

FIG. 6. The calculated and experimental [17, 18] volume thermal expansion of $M_2O$ (M=Ag, Au and Cu). The calculations for $Ag_2O$ [16] are shown here for comparison with $Au_2O$ and $Cu_2O$.

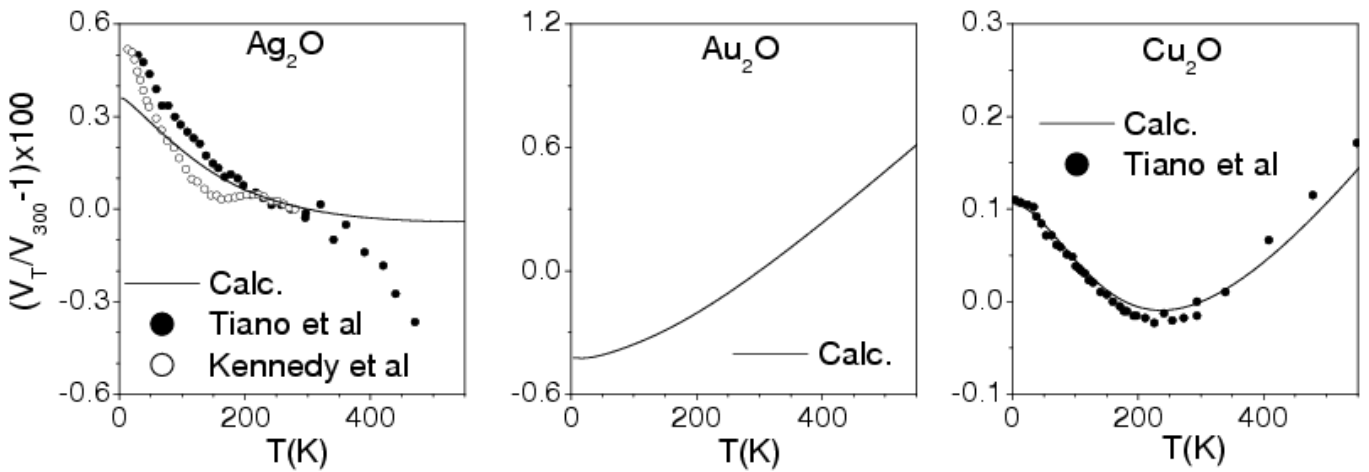

FIG. 7. (Color online) Volume thermal expansion (α) coefficient contributed from phonons of energy E.

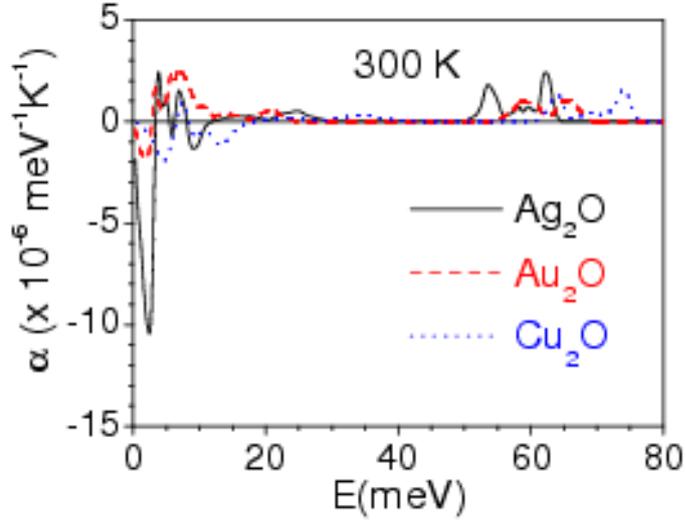

FIG. 8. (Color online) Polarization vectors of selected phonon modes in $M_2O$ (M=Ag, Au and Cu). The numbers after the wave vector (Γ, X, M and R) gives the Grüneisen parameters of $Ag_2O$, $Au_2O$ and $Cu_2O$ respectively. The lengths of arrows are related to the displacements of the atoms. Key: M, grey spheres; O, brown spheres.

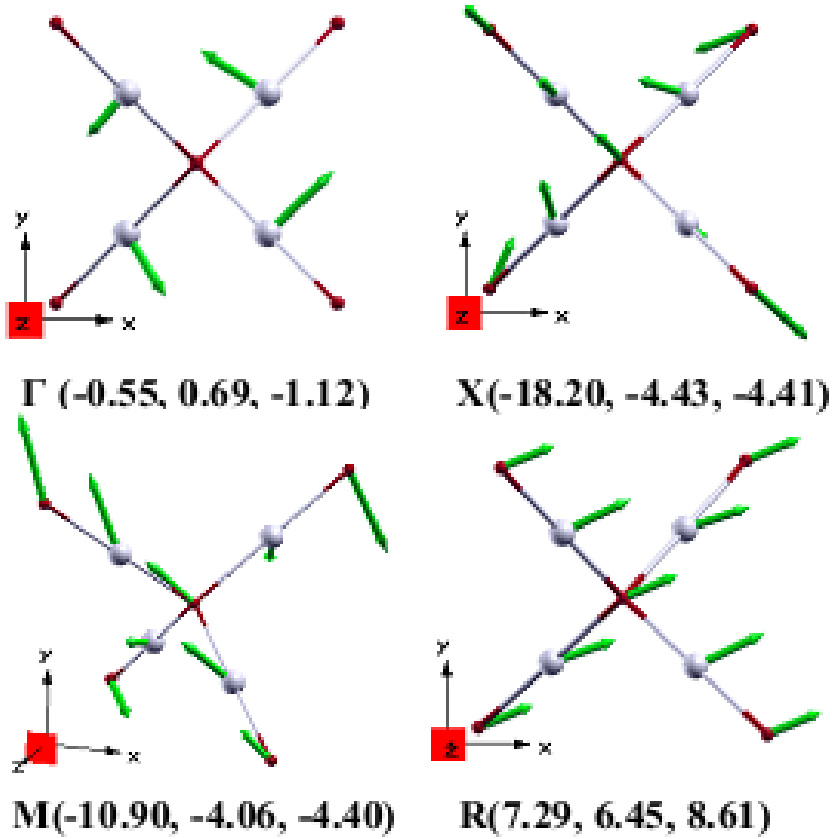

FIG. 9. (Color online) Calculated and experimental [23, 24] specific heat as a function of temperature of $M_2O$ (M=Ag, Au and Cu).

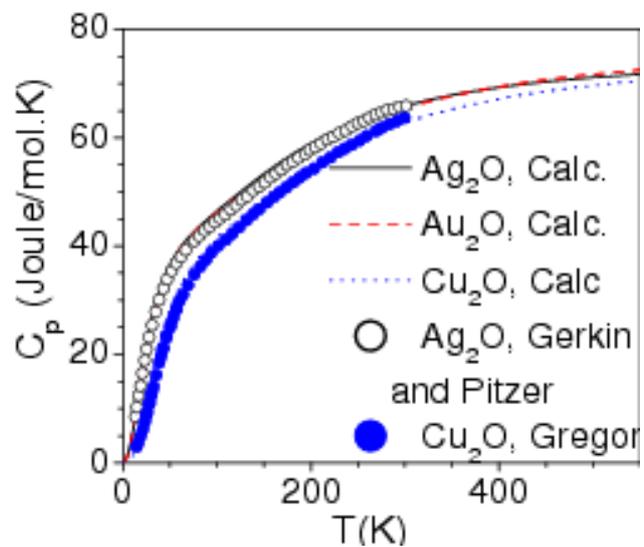

FIG. 10. (Color online) The calculated mean square amplitudes of various atoms in $M_2O$ (M=Ag, Au and Cu). The experimental data of $Cu_2O$ are from Ref. [41].

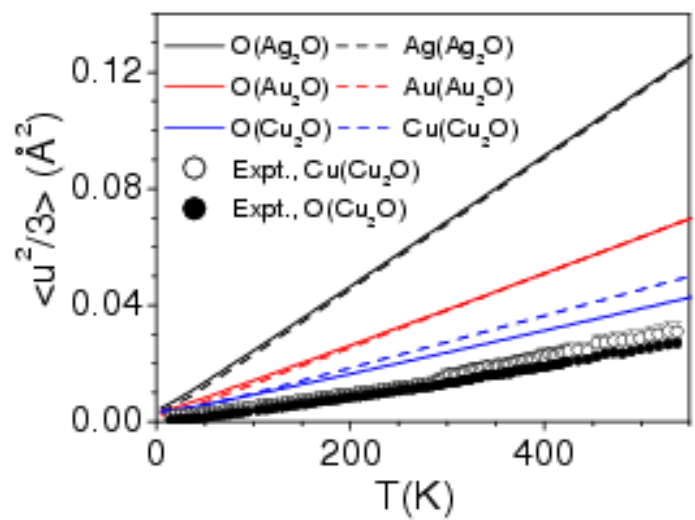

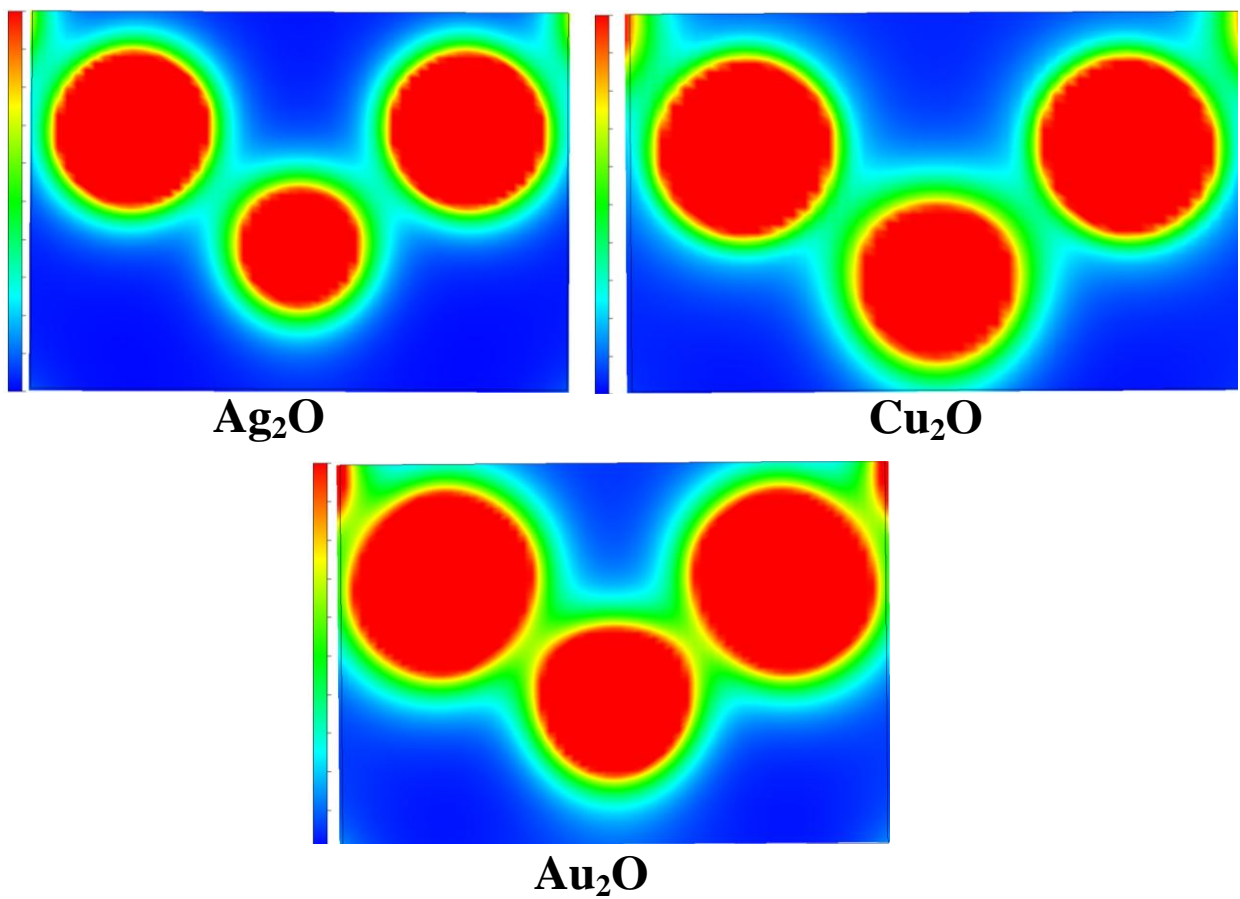

FIG. 11. (Color online) The calculated charge density for Ag$_2$O, Cu$_2$O and Cu$_2$O in (011) plane. M (Ag, Au, Cu) are the central atoms while rest are oxygen's. The b and c-axes are in the horizontal and vertical directions respectively, while a-axis is out of plane.

**FIG. 12** (Color online) The calculated potential energy profile of selected phonon modes in $M_2O$ (M=Ag, Au and Cu). Γ, X, M and R are the high symmetry points in the cubic Brillouin zone. The numbers after the wave vector give the phonon energies and Grüneisen parameters respectively.

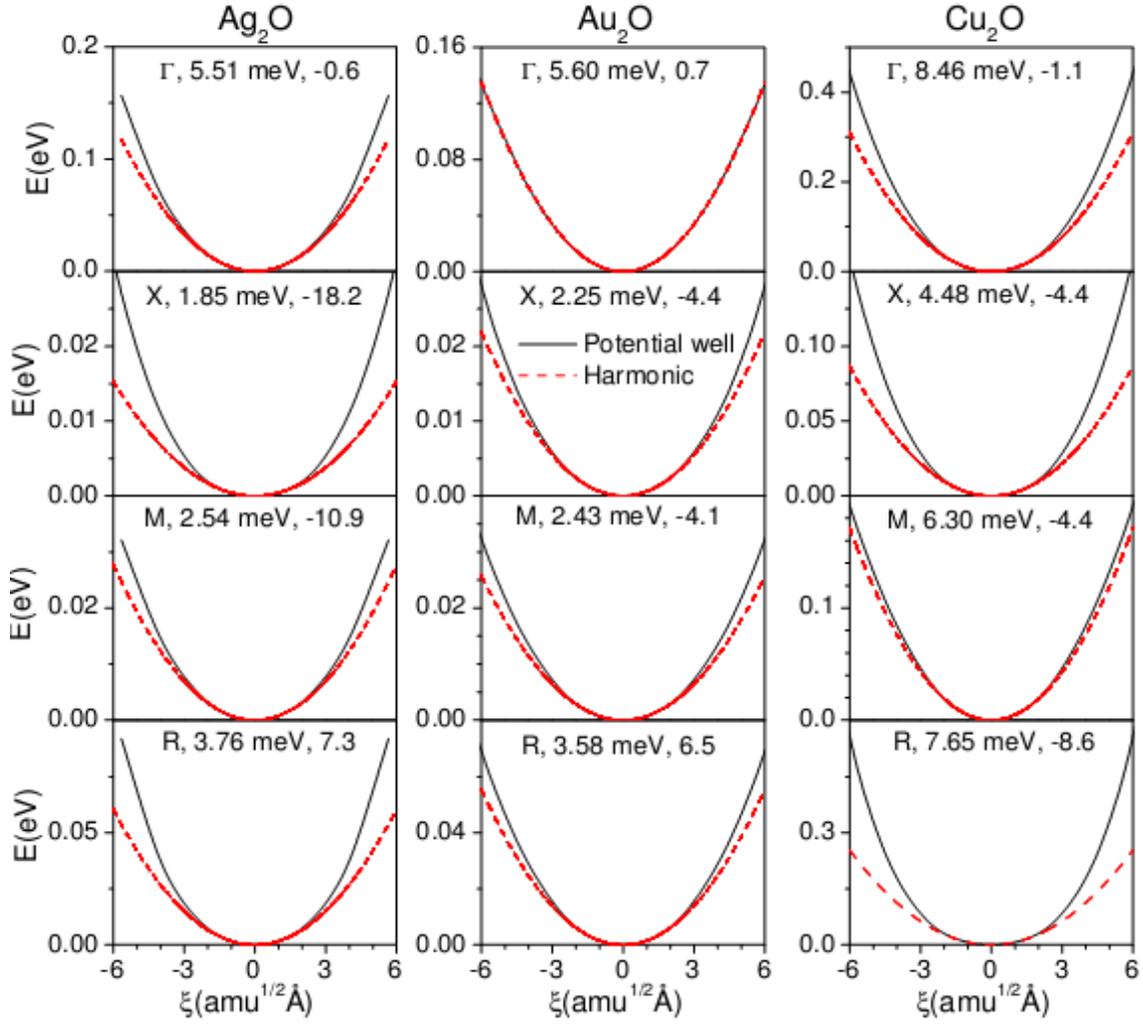

**FIG. 13** (Color online) The calculated temperature dependence of selected phonon modes in M$_2$O (M=Ag, Au and Cu). The temperature dependence of phonon modes is due to both the "implicit" as well as "explicit" anharmonicities. Γ, X, M and R are the high symmetry points in the cubic Brillouin zone. The numbers after the wave vector give the phonon energies and Grüneisen parameters respectively. Solid circles are the data taken from the experimental temperature dependence of phonon peak at 2.4 meV in the density of states [16] of Ag$_2$O, which involves average over entire Brillouin zone.

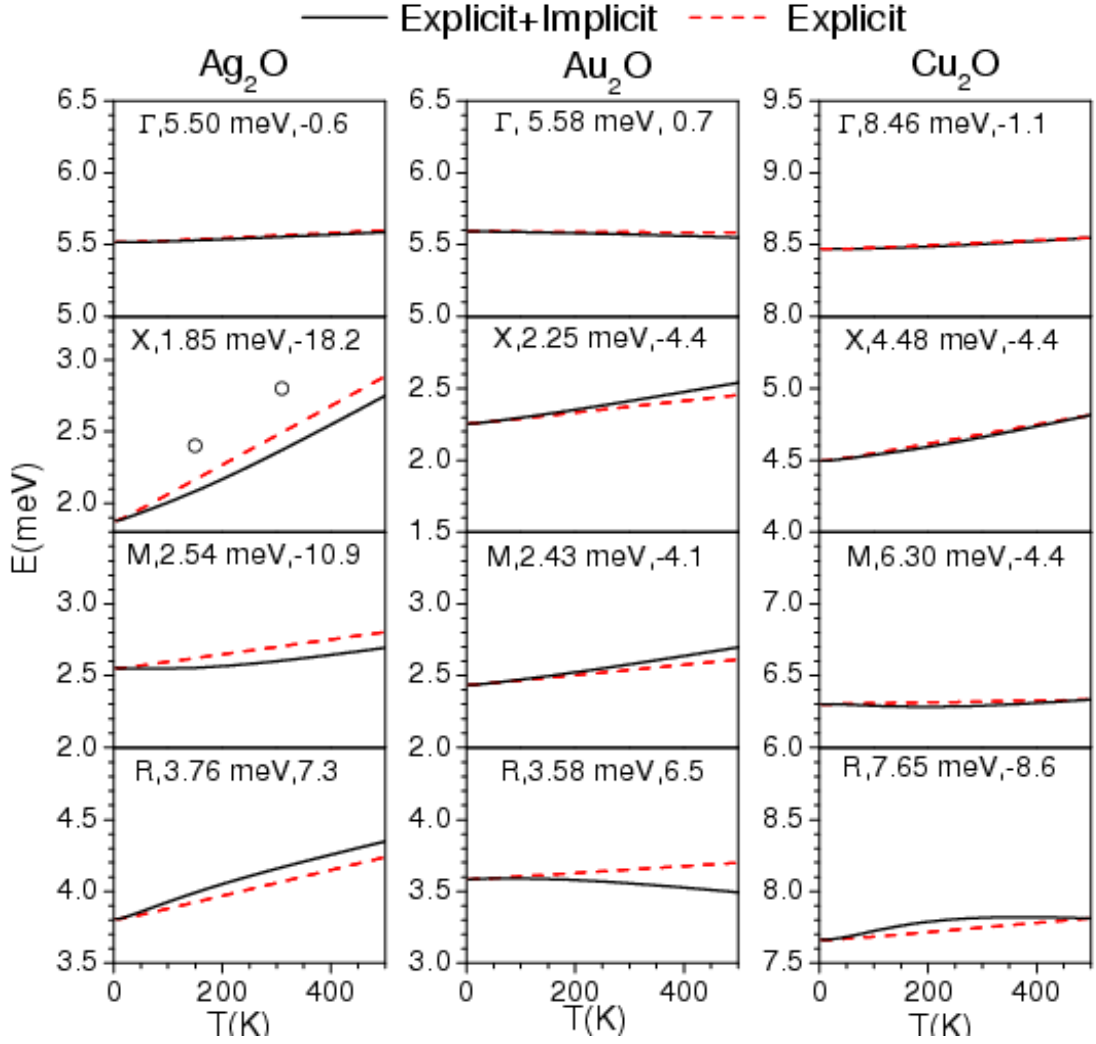

**Phonons, Nature of Bonding and Their Relation to Thermal Expansion Behavior of $M_2O$ (M=Au, Ag, Cu)**

M. K. Gupta, R. Mittal and S. L. Chaplot
*Solid State Physics Division, Bhabha Atomic Research Centre, Mumbai 400085, India*
S. Rols
*Institut Laue-Langevin, BP 156, 38042 Grenoble Cedex 9, France*

FIG. S1. The calculated phonon dispersion relation of $M_2O$ (M=Ag, Au and Cu). The Bradley-Cracknell notation is used for the high-symmetry points along which the dispersion relations are obtained. $\Gamma=(0,0,0)$; $X=(1/2,0,0)$; $M=(1/2,1/2,0)$ and $R=(1/2,1/2,1/2)$.

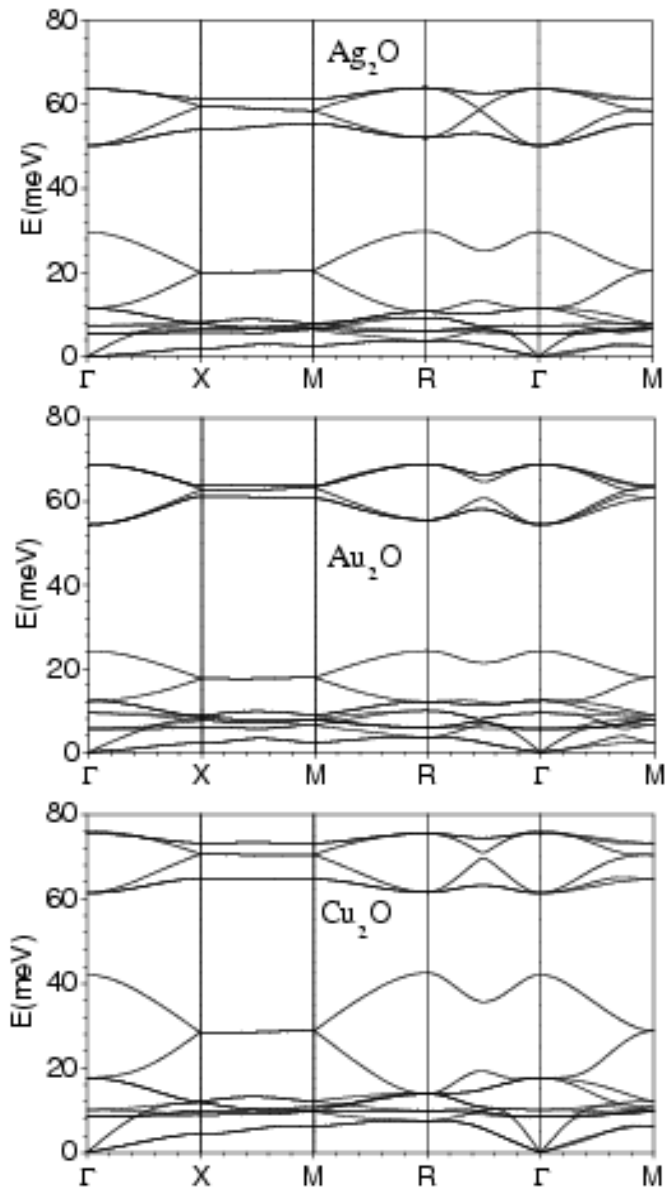

FIG. S2. (Color online) The calculated Grüneisen Parameter of $M_2O$ (M=Ag, Au and Cu).

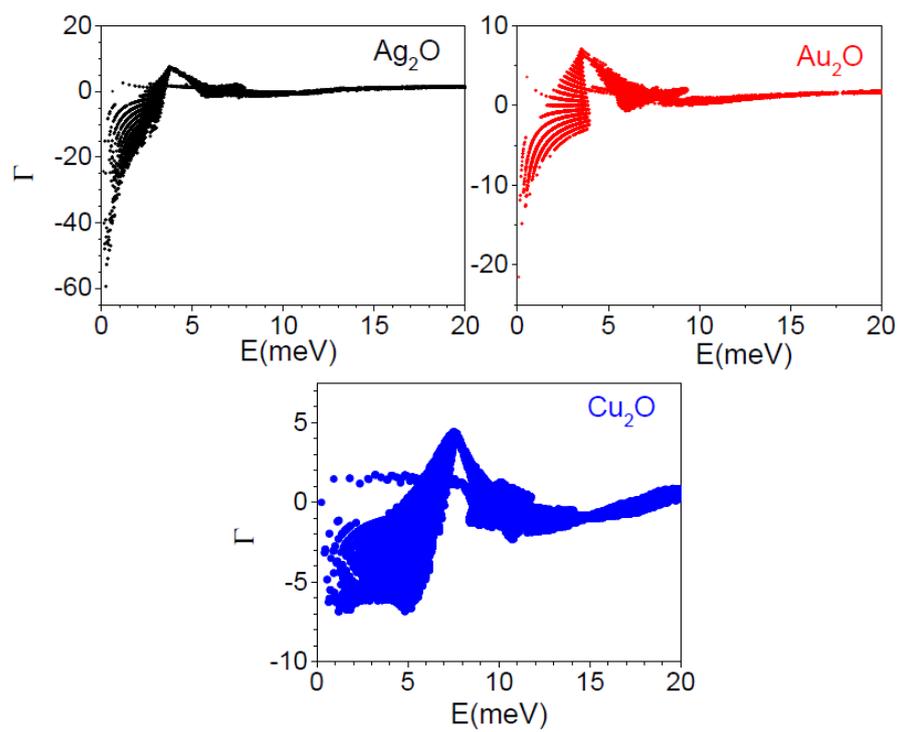